\documentclass[apl,superscriptaddress,preprint,dvipdfmx]{revtex4-1}

\newcommand {\CA}{Cd$_{\mathrm{3}}$As$_{\mathrm{2}}$}

\usepackage{graphicx}
\usepackage{bm}
\usepackage{pifont}
\usepackage{amsmath}
\usepackage{amssymb}
\usepackage{color}

\begin{document}

\title{Molecular beam epitaxy of three-dimensionally thick Dirac semimetal {\CA} films}

\author{Y. Nakazawa}
\affiliation{Department of Applied Physics and Quantum-Phase Electronics Center (QPEC), the University of Tokyo, Tokyo 113-8656, Japan}

\author{M. Uchida}
\email[Author to whom correspondence should be addressed: ]{uchida@ap.t.u-tokyo.ac.jp}
\affiliation{Department of Applied Physics and Quantum-Phase Electronics Center (QPEC), the University of Tokyo, Tokyo 113-8656, Japan}
\affiliation{PRESTO, Japan Science and Technology Agency (JST), Tokyo 102-0076, Japan}

\author{S. Nishihaya}
\affiliation{Department of Applied Physics and Quantum-Phase Electronics Center (QPEC), the University of Tokyo, Tokyo 113-8656, Japan}

\author{S. Sato}
\affiliation{Department of Applied Physics and Quantum-Phase Electronics Center (QPEC), the University of Tokyo, Tokyo 113-8656, Japan}

\author{A. Nakao}
\affiliation{RIKEN Bioengineering Laboratory, Wako 351-0198, Japan}

\author{J. Matsuno}
\affiliation{Department of Physics, Osaka University, Osaka 560-0043, Japan}

\author{M. Kawasaki}
\affiliation{Department of Applied Physics and Quantum-Phase Electronics Center (QPEC), the University of Tokyo, Tokyo 113-8656, Japan}
\affiliation{RIKEN Center for Emergent Matter Science (CEMS), Wako 351-0198, Japan}

\begin{abstract}
Rapid progress of quantum transport study in topological Dirac semimetal, including observations of quantum Hall effect in two-dimensional (2D) {\CA} samples, has uncovered even more interesting quantum transport properties in high-quality and three-dimensional (3D) samples. However, such 3D {\CA} films with low carrier density and high electron mobility have been hardly obtained. Here we report the growth and characterization of 3D thick {\CA} films adopting molecular beam epitaxy. The highest electron mobility ($\mu$ = 3 $\times$ 10$^{4}$ cm$^{2}$/Vs) among the reported film samples has been achieved at a low carrier density (\textit{n} = 5 $\times$ 10$^{16}$ cm$^{-3}$). In the magnetotransport measurement, Hall plateau-like structures are commonly observed in spite of the 3D thick films (\textit{t} = 120 nm). On the other hand, field angle dependence of the plateau-like structures and corresponding Shubunikov-de Haas oscillations rather shows a 3D feature, suggesting the appearance of unconventional magnetic orbit, also distinct from the one described by the semiclassical Weyl-orbit equation.
\end{abstract}

\maketitle
Topological Dirac semimetal (DSM) is characterized by surface Fermi arcs and bulk Dirac dispersions along any directions in three-dimensional (3D) momentum space \cite{YoungPRL2012, WangPRB2012, WangPRB2013, YangNatCommun2014, ArmitageRevModPhys2018}. As DSM is driven into other exotic topological phases including Weyl semimetal, topological insulator, and trivial insulator such as by symmetry breaking or dimensionality control \cite{WangPRB2012, WangPRB2013, LiuNatMater2014, UchidaNatComm2017, CollinsNature2018}, it has attracted both theoretical and experimental research efforts as an ideal parent material. DSM has also gathered significant interest because of its characteristic quantum transport originating from the non-trivial electronic structure; for examples, surface conduction based on Fermi-arcs \cite{WanPRB2011} and negative magnetoresistance induced by chiral anomaly \cite{SonPRB2013, BurkovPRL2014}. In paticular, the existence of novel orbital trajectories under a magnetic field, so called Weyl-orbit, which consists of both the surface Fermi-arcs and bulk chiral modes, has been predicted \cite{PotterNatComm2014, ZhangSciRep2016}.

Among the candidates of DSM, {\CA} is one of the most studied materials hosting simple band structure aroud the Dirac points \cite{NeupaneNatComm2014, LiuNatMater2014, JeonNatMater2014, BorisenkoPRL2014, NishihayaPRB2018}. So far, a number of studies have been reported on the quantum transport in {\CA} \cite{HePRL2014, LiangNatMater2015, WangNatComm2016, MollNature2016}, including observations of quantum Hall effect (QHE) \cite{UchidaNatComm2017, NishihayaSciAdv2018, ZhangNatComm2017_2, ZhangNature2018, StemmerPRL2018, LinPRL2019}. Particularly, QHE observed even in three-dimensional {\CA} nano-plates (\textit{t} = 70 $\sim$ 80 nm) \cite{ZhangNatComm2017_2, ZhangNature2018, LinPRL2019} has indicated the quantization of two-dimensional (2D) surface states to demand subsequent studies for investigating its origin. However, most of bulk {\CA} samples, synthesized by such as melt growth  \cite{NeupaneNatComm2014, LiuNatMater2014, JeonNatMater2014, ZhaoPRX2015, NarayananPRL2015} and chemical vapor deposition \cite{ZhangNatComm2017_2, ZhangNature2018, LinPRL2019}, has relatively high carrier density ($\geq$ 10$^{18}$ cm$^{-3}$). Therefore, fabrication of {\CA} films with lower carrier density and higher electron mobility has been required. In order to realize it, molecular beam epitaxy (MBE) is an useful technique, since the films can be grown by individually controlling the flux of each element. While there have been some reports on the MBE growth of {\CA} films \cite{StemmerAPLMat2016, StemmerPRB2017, StemmerPRB2018, XiuNPG2015, XiuSciRep2016, ChengNJP2016, XiuPRB2018}, the films showing quantum transport have not been thick enough to reflect the 3D nature of topological Dirac semimetal. In spite of its necessity, high-quality 3D {\CA} films have been hardly obtained mainly because of the high volatility of elements that require a low growth temperature.

Here we report MBE growth and characterization of 3D thick {\CA} films. The highest electron mobility has been achieved among the reported film samples. Even in the thickness regime of \textit{t} $>$ 100 nm, Hall plateau-like structures are observed together with Shubnikov-de Haas oscillations. Though the details of mechanisms are still unclear, the unconventional behaviors suggest the importance of coexistence of surface and bulk quantum states.

{\CA} films were grown in an EpiQuest RC1100 MBE system on single crystalline (111) CdTe substrates (see also supplemental material for details). CdTe has a zincblende structure, and the lattice mismatch between {\CA} and CdTe is 2.3 \%. The molecular beams were provided from a conventional Knudsen cell containing Cd (6N, Osaka Asahi Co.) and an MBE-Komponenten valved cracker source containing As (7N5, Furukawa Co.), respectively. The reservoir temperature of the cracker source was set to 600 $^{\circ}$C to sublimate arsenic as tetramers (As$_{4}$). The CdTe substrate was etched using 0.01 \% Br$_{2}$-methanol to remove native oxides just before loading it into the MBE chamber \cite{StemmerAPLMat2018}. Prior to the growth, the substrate was heated up to 500 $^{\circ}$C with supplying As flux. After confirming the change of the \textit{in situ} reflection high-energy electron diffraction (RHEED) pattern, from a three-dimensional transmission pattern to a streak pattern, the substrate was cooled down to the growth temperature of 200 $^{\circ}$C. The beam equivalent pressures were measured by an ionization gauge, and were set to 1.4 $\times$ 10$^{-4}$ Pa for Cd and 1.2 $\times$ 10$^{-4}$ Pa for As$_{4}$ during the co-deposition growth. This As-rich growth condition serves to reduce the carrier density of the {\CA} films, because As deficiency is a major origin of the electron carriers in {\CA}. The film thickness was set at 120 nm, and the growth rate was about 0.7 $\mathrm{\AA}$/s.

Figure 1 summarizes fundamental characterization results of a {\CA} thick film (sample A). As shown in the x-ray diffraction (XRD) $\theta$-2$\theta$ scan (Fig. 1(a)), the reflections from the \{1 1 2\} lattice planes of {\CA} are observed without any impurity phases.
The rocking curve of the (2 2 4) {\CA} film peak is shown in Fig. 1(b). The full width at half maximum (FWHM) of the film peak is 0.59$^{\circ}$, which is still broader than the bulk values \cite{HePRL2014}. An XRD reciprocal space map around the (10 2 12) {\CA} film and the (5 1 3) CdTe substrate peaks is shown in Fig. 1(c). The in-plane lattice constant along [1 $\overline{1}$ 0] and the out-of-plane lattice spacing along [1 1 2] of the {\CA} film are calculated to be 17.81 $\mathrm{\AA}$ and 7.321 $\mathrm{\AA}$, respectively, indicating that the {\CA} film is free from strain by the CdTe substrate.

A RHEED pattern taken along [1 1 $\overline{1}$] azimuth of the {\CA} film is shown in Fig. 1(d). It shows a typical streak pattern, indicative of two-dimensionally flat film surface. The corresponding atomic force microscopy (AFM) image and the thickness profile are shown in Fig. 1(e) and (f). The {\CA} film has step and terrace structure, and the step height corresponds to the lattice spacing of the {\CA} (112) planes ($\sim$ 7 $\mathrm{\AA}$). Figure 1(g) shows temperature dependence of the resistance $R_{\mathrm{xx}}$. The semiconducting temperature dependence down to 2 K also indicates the suppression of the As deficiency or electron carriers.

Figures 2(a)--(c) show out-of-plane transverse magnetoresistance $R_{\mathrm{xx}}$ and Hall resistance $R_{\mathrm{yx}}$ of typical {\CA} thick films (samples A, B, and C) measured at 2 K. The {\CA} samples were cut into 5 mm $\times$ 1 mm for transport measurement using the conventional four-terminal method. For the side electrodes flowing electric current, silver paste was attached to cover both of the entire sides and to avoid the current-jetting effect induced by inhomogeneous current flow \cite{ReisNJP2016}. Reflecting the low carrier densities as deduced from the Hall slope, Shubunikov-de Haas oscillations are  few but clearly observed for all the three samples. If an isotropic 3D Fermi surface is assumed for given carrier densities, the magnetic field where the system reaches the quantum limit is estimated at about 4, 5, and 7 T for samples A, B, and C, respectively. On the other hand, the quantum oscillations corresponding to higher filling factors of $\nu$ = 4 or 8 are observed around these fields. They are not the lowest filling factors, since the double ($\nu$ = 4, 6, $\ldots$) or four-fold ($\nu$ = 8, 12, $\ldots$) degeneracy is observed within the measurement field range.

The low-temperature electron mobility $\mu$ and carrier density \textit{n} of our thick films (red closed circles) are summarized in Fig. 2(d) together with other reported data. The mobility reaches a maximum of $\mu$ = 3 $\times$ 10$^{4}$ cm$^{2}$/Vs for sample A, while the carrier density is reduced down to \textit{n} = 5 $\times$ 10$^{16}$ cm$^{-3}$, which is one-and-a-half order of magnitude lower than the previously reported PLD-grown films (blue diamonds) \cite{UchidaNatComm2017, NakazawaSciRep2018}. The reduction of the carrier density is owing to the suppression of As deficiency by the As-rich growth condition.

To investigate the origins of the quantum oscillations in $R_{\mathrm{xx}}$ and the plateau-like structure in $R_{\mathrm{yx}}$, the field angle dependence of $R_{\mathrm{xx}}$ and $R_{\mathrm{yx}}$ was measured. Figures 3 (a) and (b) show their dependence taken for sample A, where the magnetic field is tilted from the out-of-plane direction ($\theta$ = 0$^{\circ}$) to the current direction ($\theta$ = 90$^{\circ}$). The quantum oscillations in $R_{\mathrm{xx}}$ and the plateau-like structures in $R_{\mathrm{yx}}$ remain to be resolved up to $\theta$ = 75$^{\circ}$. For clarity, $R_{\mathrm{xx}}$, $\frac{-d^{2}R_{\mathrm{xx}}}{dB^{2}}$, and $\frac{-dR_{\mathrm{yx}}}{dB}$ are compared for each field angle in Figs. 3(c)--(f). Even under rotation of $\theta$, the peak and valley positions confirmed in the derivatives do not largely shift. This field-angle-independent behavior of $R_{\mathrm{xx}}$ indicates that the 3D Fermi surface is realized in the thick {\CA} film, also consistent with thickness dependence previously studied using {\CA} films \cite{UchidaNatComm2017}. On the other hand, plateau-like structures in $R_{\mathrm{yx}}$ are indicative of a 2D electronic state (Figs. 2(a)--(c) and Fig. 3(b)). Namely, the correspondence of the oscillation phases between the derivative of the Hall resistance ($\frac{-dR_{\mathrm{yx}}}{dB}$) and the oscillation component of the magnetoresistance ($\frac{-d^{2}R_{\mathrm{xx}}}{dB^{2}}$) is a sign of the quantum Hall state. This coexistence of the 2D feature and its 3D angle dependence can not be explained either by a simple 2D nor 3D model.

Novel orbital trajectory, so called Weyl-orbit, has been theoretically predicted for topological semimetals under a magnetic field \cite{PotterNatComm2014, ZhangSciRep2016}, where the two Fermi arcs on opposite surfaces of the sample are connected by the bulk chiral mode ($N = 0$ Landau level). Quantum oscillations from the Weyl-orbit occur at the following magnetic fields,
\begin{equation}
\frac{1}{B_n}=\frac{e}{S_k} \left[2\pi(n+\gamma)\mathrm{cos}\theta-t(k_{\mathrm{W}\parallel}(\theta)+2k_{\mathrm{F}\parallel}(\theta))\right]
\end{equation}
where $n$ is the index of the Landau level, $\gamma$ is the constant phase offset, $S_{k}$ is the $k$-space area enclosed by the two Fermi arcs combined, $\theta$ is the tilting angle of the magnetic field from the surface normal. $k_{\mathrm{W}\parallel}(\theta)$ and $k_{\mathrm{F}\parallel}(\theta)$ are the field parallel components of the wave vector from +1 to -1 chirality Weyl nodes and that of the Fermi wave vector, respectively. However, the observed 3D angle dependence can not be explained even by the semiclassical Weyl-orbit picture. If the observed quantum oscillations can be described by Eq.(1), the angle-independent behavior is understood in a way that the total change in the right side of Eq.(1) is nearly independent of $\theta$. But, the second term $-t(k_{\mathrm{W}\parallel}(\theta)+2k_{\mathrm{F}\parallel}(\theta))$ does not change such that it compensates the change in the first term $2\pi(n+\gamma)\mathrm{cos}\theta$. Moreover, the right side of the Eq. (1) can even take negative values for a part of the measured range of $\theta$, for such low-carrier-density samples. To understand the characteristic quantum transport emerging in the 3D thick films, systematic control of the Fermi level such as by chemical substitution or electrostatic gating will be necessary as a future work.

In summary, 3D thick {\CA} films with low carrier density and high electron mobility have been obtained by using molecular beam epitaxy. The electron mobility of $\mu$ = 3 $\times$ 10$^{4}$ cm$^{2}$/Vs, the highest value among the reported film samples, has been achieved. The coexistence of Hall plateau-like structure and 3D field angle dependence can not be interpreted by either simple 2D or 3D model. The MBE-grown {\CA} films will be an ideal platform for further investigation of the quantum Hall state based on unconventional magnetic orbits which are also distinct from the one described by the semiclassical Weyl-orbit equation.

We acknowledge H. Ishizuka, H. Sakai, Y. Araki, K. Muraki, N. Nagaosa, and Y. Tokura for fruitful discussions. We also thank M. Tanaka and S. Ohya for technical advice about the handling of arsenides. This work was supported by JST PRESTO Grant No. JPMJPR18L2 and CREST Grant No. JPMJCR16F1, Japan, and by Grant-in-Aids for Scientific Research (B) No. JP18H01866 from MEXT, Japan.

\newpage
\begin{figure}
\begin{center}
\includegraphics*[width=16cm]{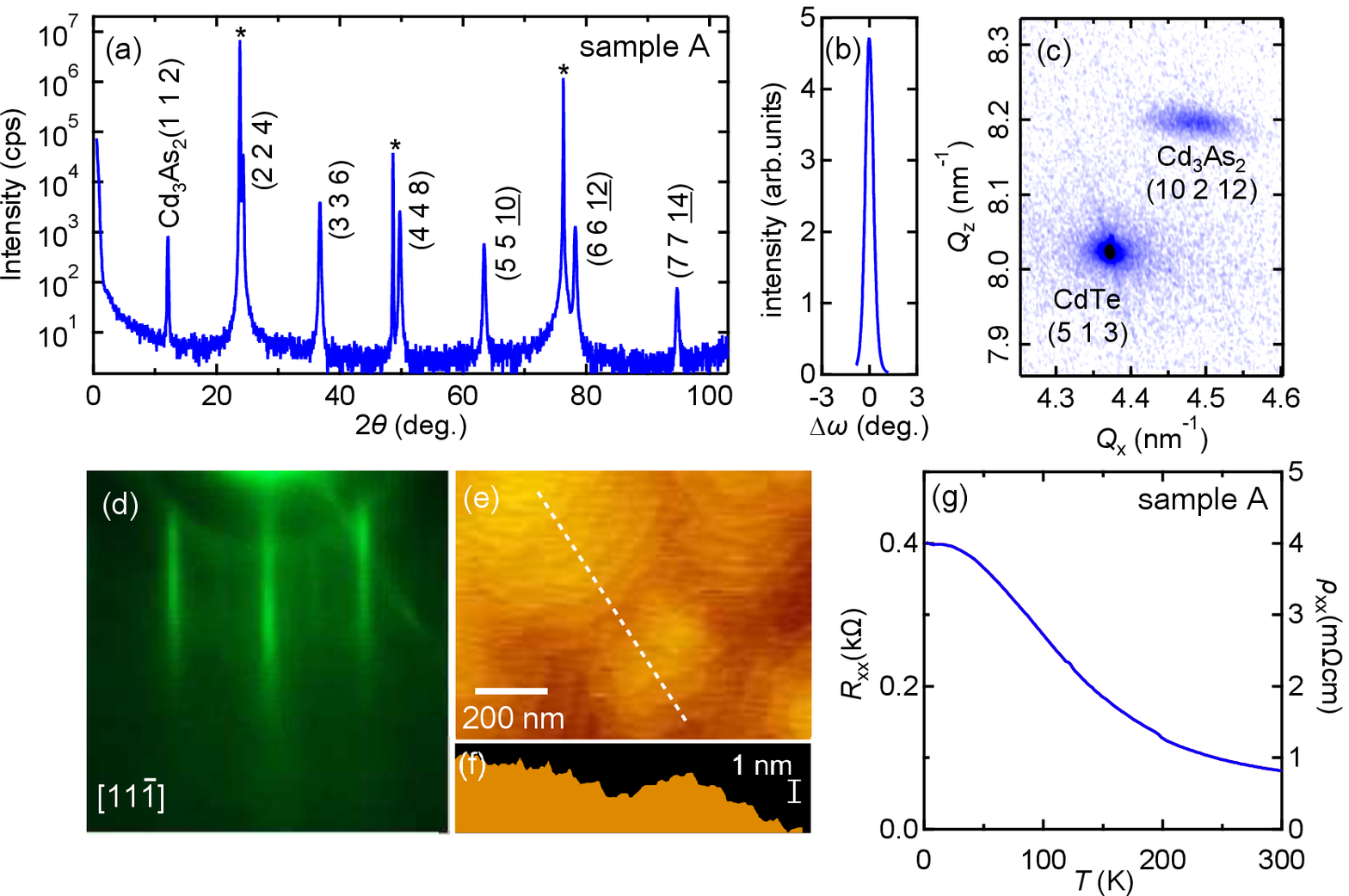}
\end{center}
\end{figure}
\noindent
FIG. 1. (a) XRD $\theta$--2$\theta$ scan of a {\CA} film (sample A) grown on a (111) CdTe substrate. (b) Rocking curve of the (224) {\CA} film peak. (c) XRD reciprocal space map around the (10 2 12) {\CA} film and the (5 1 3) CdTe substrate peaks. (d) RHEED image of the {\CA} film viewed along the [11$\overline{1}$] azimuth direction. (e) AFM image and (f) thickness profile of the film. The thickness profile is measured along the broken line in the AFM image. (g) Temperature dependence of the longitudinal resistance $R_{\mathrm{xx}}$ and resistivity $\rho_{\mathrm{xx}}$ measured for the same film.
\newpage
\begin{figure}
\begin{center}
\includegraphics*[width=14cm]{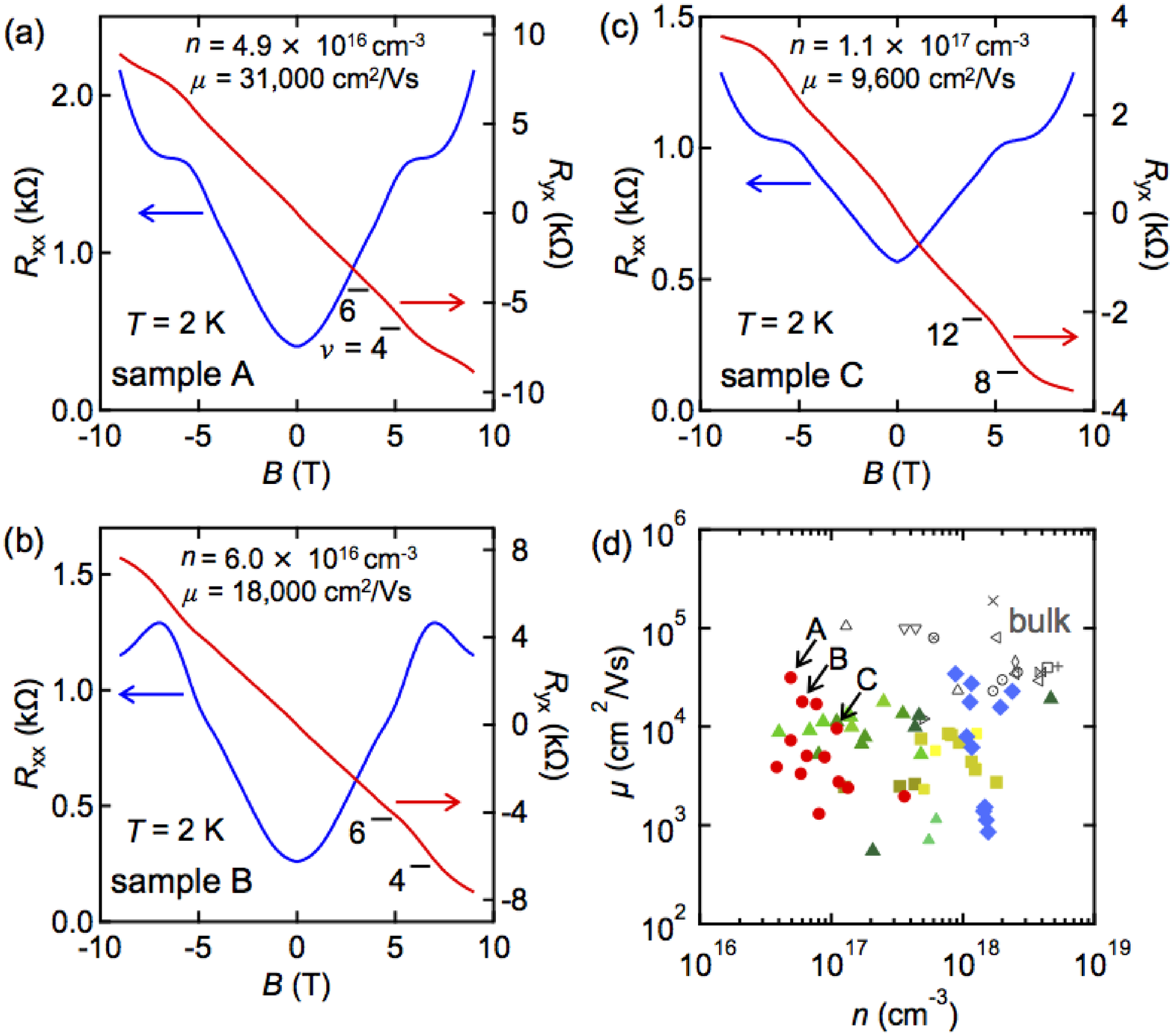}
\end{center}
\end{figure}
\noindent
FIG. 2. (a)--(c) Out-of-plane transverse magnetoresistances $R_{\mathrm{xx}}$ and Hall resistances $R_{\mathrm{yx}}$ measured for {\CA} thick films. The horizontal bars represent the quantized Hall resistances $R_{\mathrm{yx}} = -\frac{h}{\nu e^{2}}$ with the filling factors $\nu$. (d) Electron mobility $\mu$ vs. carrier density $n$ plotted for our present {\CA} films (red closed circles). For comparison, previously reported values are also plotted, including MBE-grown films at the University of California, Santa Barbara (UCSB, green triangles) \cite{StemmerAPLMat2016, StemmerPRB2017, StemmerPRL2018, StemmerAPLMat2018, StemmerPRB2018, StemmerPRMat2018}, MBE-grown films at Fudan University (yellow squares) \cite{XiuNPG2015, XiuSciRep2016, ChengNJP2016, XiuPRB2018}, our PLD-grown films \cite{UchidaNatComm2017, NakazawaSciRep2018} (blue diamonds), and bulk samples ($+$ \cite{HePRL2014}, $\lozenge$ \cite{MollNature2016}, $\triangle$ \cite{ZhangNatComm2017_2}, $\bigtriangledown$ \cite{ZhangNature2018},  $\lhd$ \cite{NarayananPRL2015}, $\square$ \cite{LiangNatMat2015},$\rhd$ \cite{ZhangNatComm2017}, $\bowtie$ \cite{Rosenberg1959}, $\times$ \cite{CaoNatComm2015}, $\odot$ \cite{Rogers1971}, and $\otimes$ \cite{WeberAPL2015}). The sample discussed in this paper are denoted as samples A, B, and C.
\newpage
\begin{figure}
\begin{center}
\includegraphics*[width=16cm]{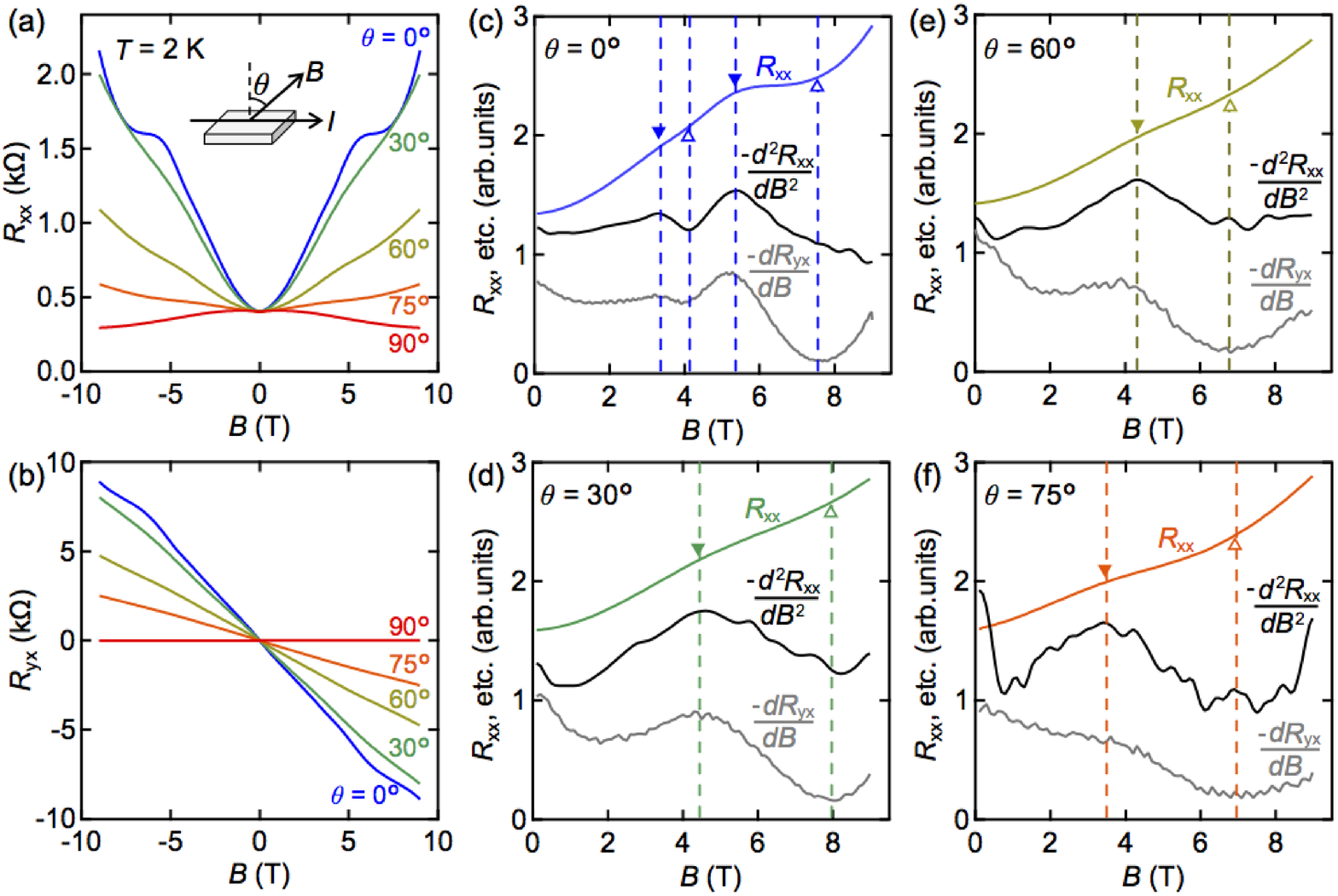}
\end{center}
\end{figure}
\noindent
FIG. 3. (a) Transverse magnetoresistance $R_{\mathrm{xx}}$ and (b) Hall resistance $R_{\mathrm{yx}}$, measured for the sample A at different field angles. (c)--(f) Field angle dependence of transverse magnetoresistance, second derivative of $R_{\mathrm{xx}}$, and first derivative of $R_{\mathrm{yx}}$.

\newpage
\begin{center}
{\large Supplementary materials for\\
\bf Molecular beam epitaxy of three-dimensionally thick\\
Dirac semimetal {\CA} films}\\
\vspace{0.15cm}
Y. Nakazawa$^1$, M. Uchida$^{1,2,*}$, S. Nishihaya$^1$, S. Sato$^1$, A. Nakao$^3$,\\
J. Matsuno$^4$, and M. Kawasaki$^{1,5}$\\
\vspace{0.15cm}
$^1${\it Department of Applied Physics and Quantum-Phase Electronics Center (QPEC),\\
the University of Tokyo, Tokyo 113-8656, Japan}\\
$^2${\it PRESTO, Japan Science and Technology Agency (JST), Tokyo 102-0076, Japan}\\
$^3${\it Institute for Innovation in International Engineering Education (IIIEE),\\
the University of Tokyo, Tokyo 113-8656, Japan}\\
$^4${\it Department of Physics, Osaka University, Osaka 560-0043, Japan}\\
$^5${\it RIKEN Center for Emergent Matter Science (CEMS), Wako 351-0198, Japan}\\
\vspace{0.15cm}
$^*$Correspondence to: E-mail: \underline{uchida@ap.t.u-tokyo.ac.jp}
\end{center}
\newpage
\noindent
{\bf Br$_2$-methanol etching of CdTe substrate}

\hspace{1cm}The CdTe substrate was etched using 0.01 \% Br$_2$-methanol just before loading it into the MBE chamber. For chemical analysis of the substrate surface, x-ray photoelectron spectroscopy (XPS) was performed on the CdTe substrate before and after the etching. Figure S1(a) shows Te 3d spectra of the CdTe substrate before (red) and after (blue) the Br$_2$-methanol etching. Before the etching, four Te 3d peaks are observed. While the two peaks at the binding energy of 573 and 584 eV are mainly assigned to Te$^{2-}$ in CdTe, the other two peaks at 577 and 587 eV are assigned to Te$^{4+}$ in TeO$_2$ [S1, S2], which is the native oxide formed on the substrate surface [S3]. These TeO$_2$ peaks are more pronounced when the incident x-ray is tilted by 30$^{\circ}$ from the plane (black), for providing higher sensitivity to the surface. After the Br$_2$-methanol etching, the peaks assigned to TeO$_2$ disappear, indicating the oxide layer is successfully removed from the CdTe substrate surface.

\vspace{1cm}
\begin{figure}[h]
\begin{center}
\includegraphics*[width=12cm]{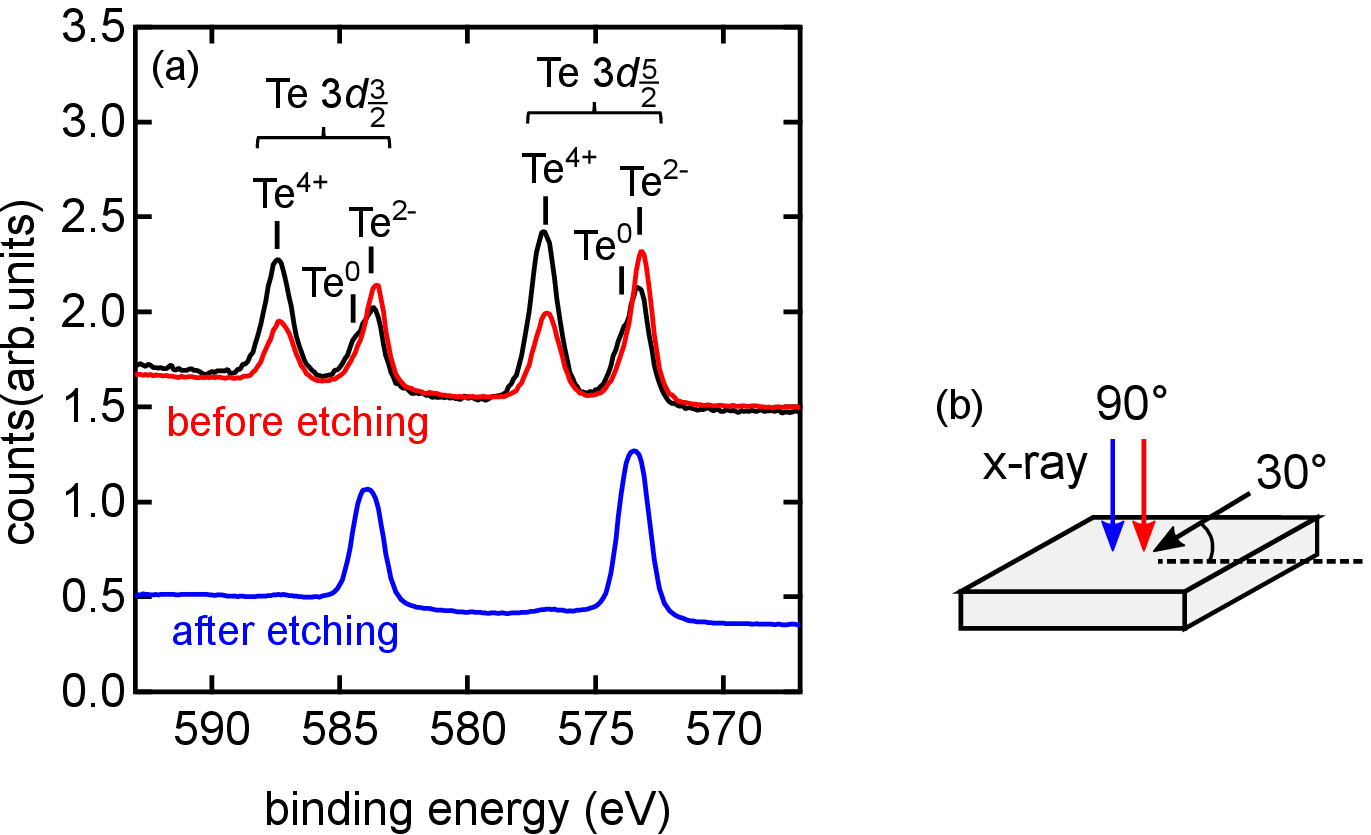}
\end{center}
\end{figure}
\noindent
FIG. S1. (a) XPS spectra of the CdTe substrate before (red) and after (blue) the etching. The spectrum taken with the incident x-ray angle of 30$^{\circ}$ is also shown (black) to show surface sensitive result. (b) Configuration of the XPS measurement.

\newpage
\noindent
{\bf Substrate annealing with supplying As$_4$ flux}

\hspace{1cm}Before the growth of {\CA} films, the CdTe substrate etched by the Br$_2$-methanol was annealed at 500 $^{\circ}$C for 5 minutes inside the MBE chamber with supplying As$_4$ flux. For clarifying the effect of the annealing process, XPS was performed for the CdTe substrates annealed with and without supplying As$_4$ flux. The two annealing sequences are shown in Figs. S2 (a) and (b). Figure S2 (c) compares As 2p spectra taken for the CdTe substrates annealed with (blue) and without (red) supplying the As$_4$ flux. The As 2p peak is observed only for the substrate annealed with the As$_4$ flux, indicating the surface modification occurs accompanied with bonding to As.

\vspace{1cm}
\begin{figure}[h]
\begin{center}
\includegraphics*[width=16cm]{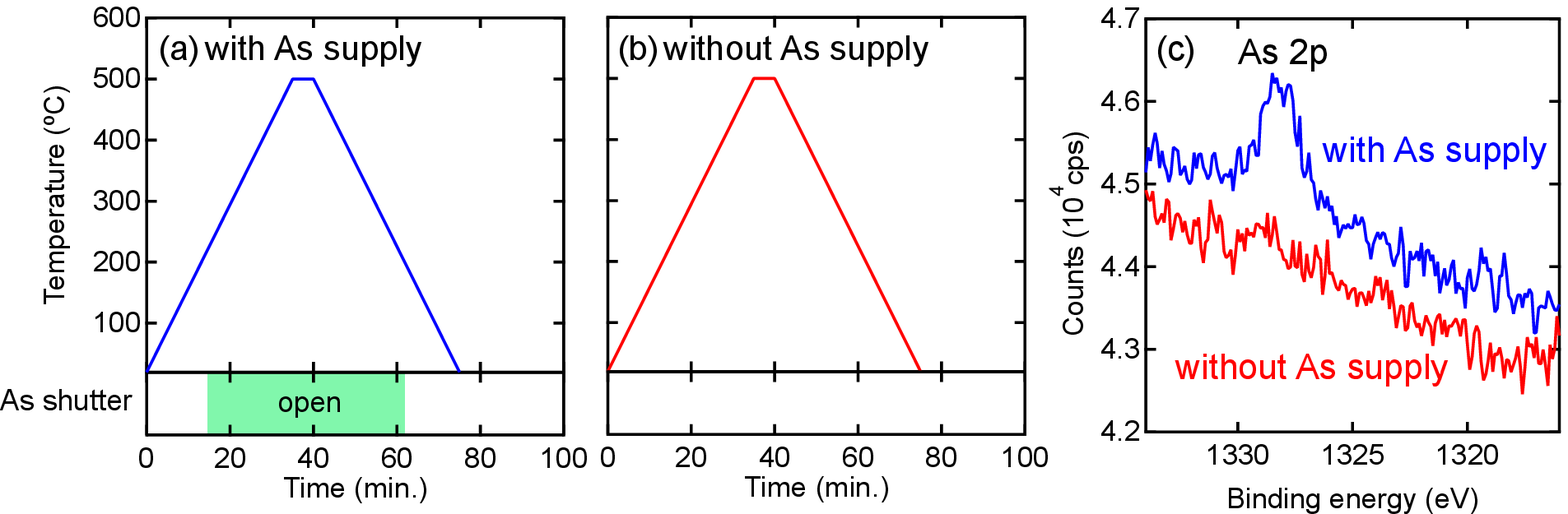}
\end{center}
\end{figure}
\noindent
FIG. S2. (a) and (b) Two annealing processes for the CdTe substrate. (c) XPS spectra taken for the CdTe substrates annealed with (blue) and without (red) supplying the As$_4$ flux.

\newpage
\noindent
{\bf Surface morphology of CdTe substrates and {\CA} films}

\hspace{1cm}Figure S3 shows the atomic force microscopy (AFM) images with thickness profiles and reflection high-energy electron diffraction (RHEED) patterns taken at representative stages during the sequence of the {\CA} film growth. Before the Br$_2$-methanol etching, particle-like structures are found on the CdTe substrate and RHEED shows a typical three-dimensional spotty pattern with low intensity (Figs. S3 (b) and (d)). After the Br$_2$-methanol etching, the particle-like structures are removed and the spotty RHEED pattern becomes clearer (Figs. S3 (e) and (g)). The RHEED pattern changes to the two-dimensional streak pattern during the annealing at 500 $^{\circ}$C with supplying the As$_4$ flux, indicating the surface modification occurs (Figs. S3(h) and (j)). After the film growth, RHEED pattern becomes sharp streaks (Fig. S3 (m)) and the AFM image (Fig. S3 (k)) is free from particles and shows step and terrace structure, as shown in Fig. 1(e) in the main text.

\newpage
\begin{figure}[t]
\begin{center}
\includegraphics*[width=16cm]{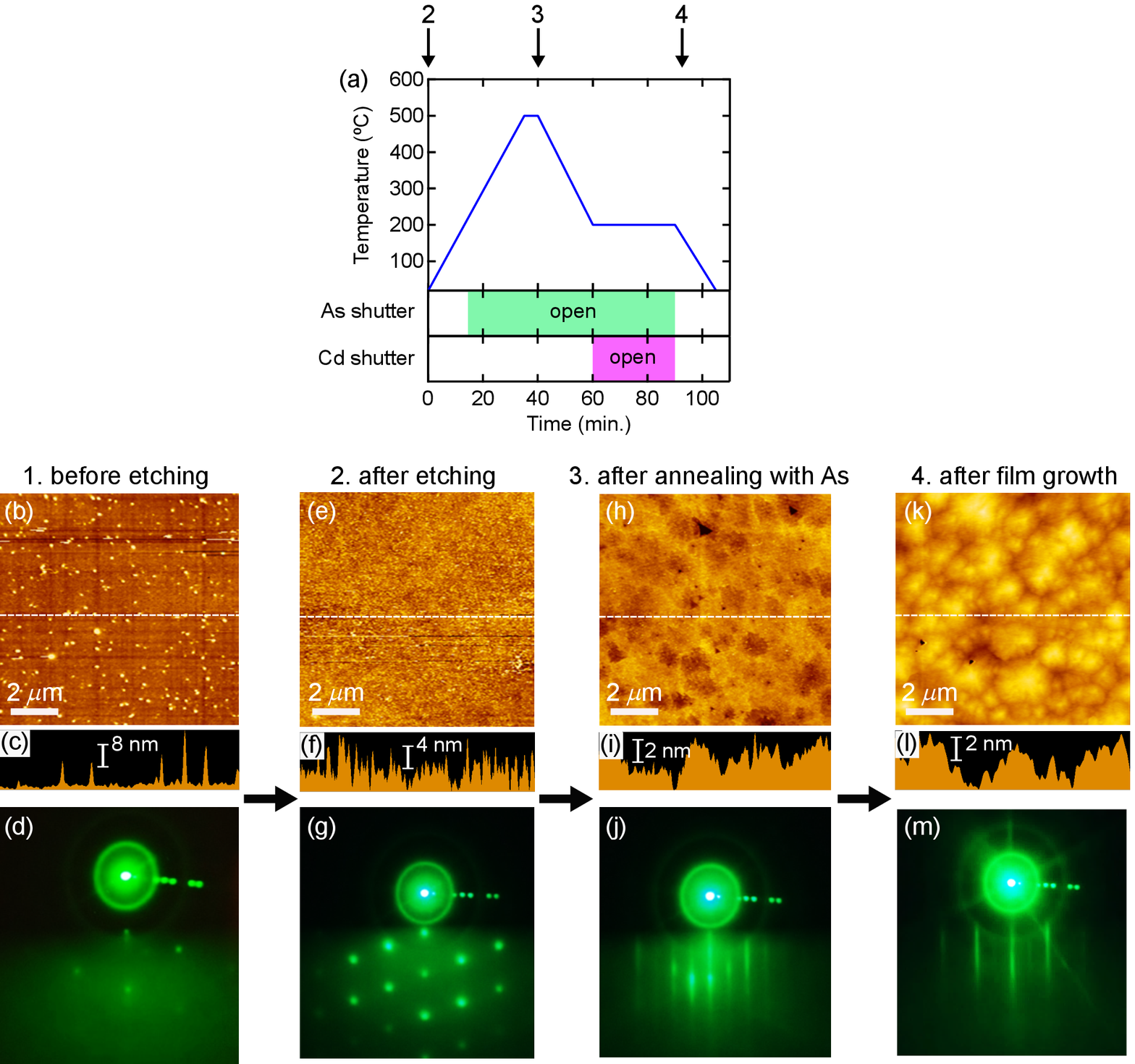}
\end{center}
\end{figure}
\noindent
FIG. S3. (a) Sequence of the {\CA} film growth. (b)-(m) AFM images, thickness profiles along the broken line, and RHEED patterns taken at representative stages indicated in (a). ((b)-(d) before the Br$_2$-methanol etching (e)-(g) after the Br$_2$-methanol etching (h)-(j) after annealing at 500 $^{\circ}$C with supplying As$_4$ flux (k)-(m) after the film growth).

\newpage
\noindent
{\large\bf Supplementary references}

\noindent
[S1] J. F. Moulder, W. F. Stickle, P. E. Sobol, and K. D. Bomben, {\it Handbook of X-Ray Photoelectron Spectroscopy} (Physical Electronics Division, Perkin-Elmer Corporation, Waltham, 1992).

\noindent
[S2] I. M. Kotina, L. M. Tukhkonen, G. V. Patsekina, A. V. Shchukarev,and G. M. Gusinskii, Semicond. Sci. Technol. {\bf 13}, 890 (1998).

\noindent
[S3] R. Triboulet and P. Siffert, {\it CdTe and Related Compounds; Physics, Defects, Hetero- and Nano-structures, Crystal Growth, Surfaces and Applications} (Elsevier, Amsterdam, 2010).

\newpage
\end{document}